\documentclass[12pt]{article}
\usepackage{a4,amsmath,amssymb,cite,graphicx}
\def\Bbb{\mathbb}
\def\BZ{\Bbb Z} 
 
\def\BH{\mathbb{H}}

\catcode`@=11 \@addtoreset{equation}{section} \catcode`@=12

\begin{document}
\bibliographystyle{utphys}
\begin{titlepage}
\renewcommand{\thefootnote}{\fnsymbol{footnote}}
\noindent
{\tt HRI/ST/1012}\hfill

\begin{center}
\large{\bf  BKM Lie superalgebra for the $Z_5$-orbifolded CHL string}
\end{center} 
\bigskip 
\begin{center}

K. Gopala Krishna\footnote{\texttt{krishna@mpim-bonn.mpg.de}} \\

\textit{Max-Planck-Institut f$\ddot{u}$r Mathematik, \\
Vivatsgasse 7 \\
53111 Bonn, Germany}
\end{center}

\begin{abstract}
We study the $\mathbb Z_5$-orbifolding of the CHL string theory by explicitly constructing the modular form $\widetilde{\Phi}_2$ generating the degeneracies of the $\tfrac14$-BPS states in the theory. Since the additive seed for the sum form is a weak Jacobi form in this case, a mismatch is found between the modular forms generated from the additive lift and the product form derived from threshold corrections. We also construct the BKM Lie superalgebra, $\widetilde{\mathcal G}_5$, corresponding to the modular form $\widetilde{\Delta}_1 (\mathbb Z) = \widetilde{\Phi}_2 (\mathbb Z)^{1/2}$ which happens to be a hyperbolic algebra. This is the first occurrence of a hyperbolic BKM Lie superalgebra. We also study the walls of marginal stability of this theory in detail, and extend the arithmetic structure found by Cheng and Dabholkar for the $N=1,2,3$ orbifoldings to the $N=4,5$ and $6$ models, all of which have an infinite number of walls in the fundamental domain. We find that analogous to the Stern-Brocot tree, which generated the intercepts of the walls on the real line, the intercepts for the $N > 3$ cases are generated by linear recurrence relations. Using the correspondence between the walls of marginal stability and the walls of the Weyl chamber of the corresponding BKM Lie superalgebra, we propose the Cartan matrices for the BKM Lie superalgebras corresponding to the $N=5$ and $6$ models. 
\end{abstract}
\end{titlepage}
\setcounter{footnote}{0}
\section{Introduction}
There has recently been renewed interest, and considerable activity, in understanding the area phrased by Harvey and Moore as \textit{`the algebra of BPS states'}\cite{Harvey:1996gc, Harvey:1995fq}. Our interest in the present paper is in constructing the \textit{`algebra'} in the context of the CHL models. Harvey and Moore considered BPS states in string theory with $\mathcal N =2$ spacetime supersymmetry and showed that the threshold corrections in the $\mathcal N = 2$ heterotic string compactifications are determined in terms of the spectrum of the BPS states \cite{Harvey:1995fq}. They found a connection between the threshold correction integrals and infinite product representations of holomorphic automorphic forms studied by Borcherds earlier\cite{Borcherds:1995Os+2, Borcherds:1995Os2GKM}. Later Borcherds, using the regularization of the integral given by Harvey and Moore, constructed a generalization of the Rankin-Selberg method to obtain automorphic forms on Grassmannians which have singularities along sub Grassmannians\cite{Borcherds:1998Aut}. In particular, for the case of unimodular lattices $\mathbb R^{2,s}$, the results of\cite{Borcherds:1995Os+2, Borcherds:1995Os2GKM} for families of holomorphic automorphic forms could be re-derived using the general method of \cite{Borcherds:1998Aut} in much simpler fashion. 

In the context of counting $\tfrac14$-BPS states in string theory, these very automorpic forms appear as generating functions of dyonic degeneracies. The degeneracies of BPS states preserving one fourth supersymmetry in a class of $\mathcal N=4$ supersymmetric string theories in four dimensions are found to be generated by the modular forms obtained from the generalized theta correspondence, while the degeneracies of those preserving half the supersymmetry are given by cusp forms of $\Gamma_0(N)$. The generalized theta correspondence of Borcherds, in this context, gives the threshold integral starting from the elliptic genera of $K3$. In the prototypical example of the type II string theory compactified on $K3 \times T^2$ or equivalently, the heterotic string compactified on $T^6$, the degeneracies of the $\tfrac12$-BPS states are generated by the weight $12$ cusp form for $\Gamma$, $\eta(\tau)^{24}$, while the degeneracies of the $\tfrac14$-BPS states are generated by the Igusa cusp form, $\Phi_{10} (\mathbf Z)$, of weight $10$ for $Sp(2, \mathbb Z)$. More generally, when this theory is considered with the compactification space orbifolded by a $\mathbb Z_N$-group whose action preserves the $\mathcal N=4$ supersymmetry, the degeneracies of the  $\tfrac12$-BPS states are found to be generated by genus-one cusp forms of $\Gamma_0(N)$, while the degeneracies of the $\tfrac14$-BPS states are generated by genus-two Siegel modular forms of suitable level $N$ subgroups of $Sp(2, \mathbb Z)$. 

The fact that the degeneracies of the BPS states are generated by modular forms obtained from the generalized theta correspondence gives rise to the possibility that they may have a BKM Lie superalgebra associated with them since such structures were found in the construction of Borcherds. Indeed, the square root of the weight $10$ Igusa cusp form of $Sp(2, \mathbb Z)$ also occurs as the denominator identity of a rank $3$ Borcherds-Kac-Moody (BKM) Lie superalgebra, denoted $\mathcal G_1$, constructed by Gritsenko and Nikulin\cite{Nikulin:1995}. Along similar lines, BKM Lie superalgebras (all, like $\mathcal G_1$, of rank $3$) have been found to exist corresponding to the genus-two Siegel modular forms occurring in a family of four-dimensional $\mathcal N =4$ supersymmetric string theories, known as the CHL strings\cite{Gritsenko:2002, Cheng:2008kt, Govindarajan:2009qt, Govindarajan:2010ge}. The fact that for the orbifolded theories there exists more than one cusp, and hence more than one infinite product expansion of the modular form at each of the cusps, leads to the existence of more than one familiy of BKM Lie superalgebras associated with the orbifolded models. This is similar to the case of the fake monster superalgebra where the BKM Lie superalgebras associated to the two cusps at level $1$ and $2$ are distinct and very different from each other although the denominator identities of the two algebras are the expansions of the same function, albeit about different cusps. For the CHL family, this leads to two different families of BKM Lie superalgebras with very distinct structures and properties. A list of all the modular forms arising in the CHL models that have a corresponding algebra structure can be found in the ``periodic table" of BKM Lie superalgebras listed by Govindarajan\cite{Govindarajan:2010ge}. 

The main challenge is to make the algebras constructed relevant to the physical theory from which they arise for which one needs to find relations between the two that can help study one from the other. In what is hopefully the first step towards setting up a dictionary between the algebraic and physical sides of the BPS state counting, it was observed by Cheng and Verlinde that the walls of marginal stability of the $\tfrac14$-BPS states are in one-to-one correspondence with the walls of the Weyl chambers of the corresponding BKM Lie superalgebras\cite{Cheng:2008fc}. This correspondence is an indication that the BKM Lie superalgebras are not mere academic constructions from the modular forms, but should actually be related to physical aspects of the CHL theory. 

That the counting of BPS states should have an algebraic structure associated to them is both intriguing and promising for the theory. Intriguing, for the reasons of their origin are not clearly known and are not such as to be foreseen at the level of the action of the theory, and promising for the possibilities it presents to know more about the microscopic side and also for unearthing other deeper structures of the theory. Already newer and unexpected relations are being uncovered based on these ideas, like the moonshine conjecture for $M_{24}$\cite{Cheng:2010k3, Eguchi:2010mo} etc.. 

In this work we study the $\mathbb Z_5$-orbifolded CHL theory and the BKM Lie superalgebra structure arising in it. We also extend Sen's study of the walls of marginal stability\cite{Sen:2007vb} of the $\tfrac14$-BPS states in the CHL models.  

\subsection*{Organization of the paper}

The organization of the paper is as follows. In section $2$, we discuss the details of the counting of BPS states by providing a brief introduction to the setting of the problem. In section $3$, we provide the details of the construction of the relevant modular forms that are the generating functions of the half and quarter BPS states in the CHL models in general, and in the $\mathbb Z_5$-orbifold in particular. In section $4$, we discuss the BKM Lie superalgebras arising from these modular forms. We show that the `square root' of the modular forms constructed in section $3$ appear as the denominator formulae of BKM Lie superalgebras. We show that the BKM Lie superalgebra for the modular form $\widetilde{\Delta}_1(\mathbb Z) = \widetilde{\Phi}_2(\mathbb Z)^{1/2}$ exists and is a hyperbolic one with an infinite number of real simple roots. The BKM Lie superalgebra has two sets of roots which are copies of one another. In section $5$, we study the walls of marginal stability of the $\tfrac14$-BPS states in the CHL models, and study the $N=5$ case in section $5.1$. In section $5.2$, we study the BKM Lie superalgebra $\mathcal{\widetilde G}_5$ in relation to the correspondence between the walls of marginal stability of the $\tfrac14$-BPS states and the walls of the Weyl chamber of $\widetilde{\mathcal G}_5$. We use this correspondence to label the roots of the algebra from the corresponding labeling of the walls, and from this derive the Cartan matrix of the algebra $\widetilde{\mathcal G}_5$. We also show the invariance of the modular form under the Weyl group of the algebra. In section $6$, we study the arithmetic structure in the walls of marginal stability of the $\tfrac14$-BPS states in the CHL models. This is similar to that found by Cheng and Dabholkar in \cite{Cheng:2008kt} for the $N=1,2,3$ models, where the roots were given by the intercepts on the real line in the Stern-Brocot tree. We find that the walls are generated by linear recurrence relations in $N=4,5,6$ and possibly $7$, replacing the Stern-Brocot tree which generated the walls in the $N=1,2,3$ cases.

\section{The counting of BPS states}

Although initial progress on microscopic couting goes back some time before it, the starting point for us, in the direction we are interested in, will be the work of Dijkgraaf, Verlinde and Verlinde. More than a decade ago, Dijkgraaf, Verlinde and Verlinde (DVV) proposed a microscopic index formula for counting the degeneracy of $\tfrac14$-BPS dyons in heterotic string theory compactified on a six-torus\cite{Dijkgraaf:1996it}. It is known that the degeneracy of electric $\tfrac12$-BPS states, which can be understood as the states of the heterotic string with the supersymmetric sector in the ground state, is generated by $1/ \eta(\tau)^{24}$. There is an $SL(2, \mathbb Z)$ electric-magnetic duality symmetry, which implies that the magnetic $\tfrac12$-BPS states, which necessarily arise non-perturbatively as solitonic states, are also generated by the same modular form. To generalize this to the dyonic states, DVV's basic idea was to think of a dyonic state carrying electric and magnetic charges as a bound state of an electric heterotic state with a dual magnetic heterotic state and using this picture to construct a modular form which counts the degeneracy of the $\tfrac14$-BPS states as a generalization of the one that counts the degeneracies of the $\tfrac12$-BPS states. The appropriate generalization turned out to be a genus-two Siegel modular form of weight $10$, which is the unique cusp form for the modular group $Sp(2, \mathbb Z)$. Intuitively, the $2\times 2$ period matrix of the Siegel modular form can be thought of as parameterizing a genus-two Riemann surface made up of two genus-one surfaces whose moduli come from the $\tfrac12$-BPS generating functions. The degeneracy, $D(n,\ell,m)$, of a dyonic state carrying charges $(n,\ell,m)=(\frac12\mathbf{q_e}^2, \mathbf{q_e}\cdot \mathbf{q_m}, \frac12\mathbf{q_m}^2)$ are generated by the Siegel modular form as 
 \begin{equation}
\label{14BPSdegeneracies}
\frac{64}{\Phi_{10}(\mathbf Z)} = \sum_{(n,\ell,m)} D(n,\ell,m) \ q^n r^\ell s^m\ ,
\end{equation}
where $\mathbf{Z}\in \BH_2$, the Siegel upper-half space and $(\frac12\mathbf{q_e}^2, \mathbf{q_e}\cdot \mathbf{q_m}, \frac12\mathbf{q_m}^2)$ are the T-duality invariant combinations of electric and magnetic charges. The above construction has since then been extended to other settings in four-dimensions with $\mathcal N=4$ supersymmetry, notably  the family of CHL strings and type II compactifications\cite{Jatkar:2005bh, David:2006ru}. We study a particular case of the CHL orbifoldings, the $\mathbb Z_5$-orbifolding, in this work. 

The CHL orbifolds arise as a family of asymmetric $\mathbb{Z}_N$-orbifolds of the heterotic string preserving the $\mathcal{N} =4$ supersymmetry of the unorbifolded theory. Four-dimensional compactification of string theory with $\mathcal{N} =4$ supersymmetry has three perturbative formulations in terms of toroidally compactified heterotic string and type IIA/B string theory compactified on $K3 \times T^2$. Consider the heterotic string compactified on a six-torus, $T^6 = T^4 \times S^1 \times \widetilde{S}^1$. The generator of the $\BZ_N$-orbifolding acts by a $1/N$ shift along the circle $S^1$ and a simultaneous $\BZ_N$-involution of the Narain lattice, $\Gamma^{20,4}$, associated with the $T^4$. On the dual type II side, the orbifolding action corresponds to an order $N$ automorphism of $K3$ which preserves the holomorphic two-form, together with a $1/N$ translation along one of the $S^1$ of the $T^2$. 

Upon orbifolding, the $\mathbb Z_N$ action gives rise to twisted states in the theory and the vector multiplet moduli space for the theory gets modified to
\begin{equation}
\big(\Gamma_1(N)\times SO(6,m;\BZ)\big)\bigg\backslash\! \left(\frac{SL(2)}{U(1)} \times \frac{SO(6,m)}{SO(6)\times SO(m)}\right).
\end{equation}
The group $SO(6,m;\BZ)$ is the T-duality symmetry group where $m= [48/(N+1)] -2$ and $\Gamma_1(N)\subset PSL(2,\BZ)$ is the S-duality symmetry group that is manifest in the equations of motion and is compatible with the charge quantization. Notice that the orbifolding breaks the S-duality group from $SL(2, \mathbb Z)$ to the subgroup $\Gamma_1(N)$. 

Extending the counting of states from the toroidally compactified heterotic strings as given by DVV, to the CHL orbifolds, Jatkar and Sen constructed two families of genus-two Siegel modular forms, $\widetilde{\Phi}_k(\mathbf{Z})$ and $\Phi_k (\mathbf Z)$, for the CHL models with $\mathbb Z_N$-orbifoldings, when $N$ is prime\cite{Jatkar:2005bh}. The weight, $k$, of the modular form is related to the orbifolding group $\BZ_N$ by $(k+2) = 24/(N+1)$ (where $N$ is prime and $(N+1)|24$). The appearance of two different families is due to the fact that for the orbifolded theories the modular group $Sp (2, \mathbb Z)$ is broken down to a smaller subgroup and hence there is more than one cusp. The expansion about each cusp gives a different family of algebras. The family of modular forms, $\widetilde{\Phi}_k(\mathbf{Z})$, are the generating functions for the degeneracies of $\tfrac14$-BPS states in the CHL models, while the family $\Phi_k (\mathbf Z)$ generates the degeneracies for the twisted dyonic states\cite{Sen:2010tw, Govindarajan:2010ge}. From these modular forms, the dyon degeneracy is given by a relation of the type \eqref{14BPSdegeneracies}. Govindarajan and Krishna\cite{Govindarajan:2009qt} extended this work by constructing the modular forms generating dyon degeneracies for composite $N$. In particular, the case of $\mathbb Z_4$-orbifolding was explicitly worked out and the corresponding modular forms, $\widetilde{\Phi}_3 (\mathbf Z)$ and $\Phi_3 (\mathbf Z)$, were constructed in\cite{Govindarajan:2009qt}. Merging the two families and extending the above constructions, Govindarajan recently constructed the family of modular forms, $\Phi_k^{(N,M)}(\mathbf Z)$, which generate the degeneracies of the $\mathbb Z_M$-twisted dyonic states in the CHL $\mathbb Z_N$-orbifolds\cite{Govindarajan:2010ge}. They also incorporate the two families, $\widetilde{\Phi}_k(\mathbf{Z})$ and $\Phi_k (\mathbf Z)$, as particular cases and correspond to $\Phi_k^{(1,M)}(\mathbf Z)$ and $\Phi_k^{(N,1)}(\mathbf Z)$, respectively, in the list. This completes the general construction of the modular forms generating the dyonic degeneracies in the CHL theories. 

Simultaneously, as one constructs the modular forms, one is also interested in exploring if a BKM Lie superalgebra structure, such as was seen for $\Phi_{10} (\mathbf Z)$, exists for the other modular forms constructed in the context of the CHL strings. BKM Lie superalgebras for the $N=1,2,3$ and $4$ theories have been constructed and studied in\cite{Cheng:2008kt, Govindarajan:2008vi, Govindarajan:2009qt, Govindarajan:2010ge}. However, for $N>5$ in the $\mathbb Z_N$-orbifolded CHL theories, the algebraic side of the theory has not been studied in detail. In this work we propose to study the $N=5$ orbifolding of the CHL theory and construct the algebraic structure corresponding to it. We also provide support for the existence of similar algebras for $N=6$. We begin by first discussing the construction of the modular forms that generate the degeneracies of the $\tfrac14$-BPS states in the theory. 

\section{The modular forms for the $\mathbb Z_5$-orbifolded CHL theory}

\subsection{Counting $\frac12$-BPS states and the additive lift}

In DVV's construction the counting of $\frac12$-BPS states in the toroidally compactified heterotic string formed the starting point for the construction of the genus-two Siegel modular form generating the degeneracies of the $\frac14$-BPS states. The counting of the degeneracy of $\tfrac12$-BPS states of a given electric charge is mapped to the counting of states of the heterotic string with the supersymmetric right-movers in the ground state\cite{Sen:2005ch,Dabholkar:2005dt,Dabholkar:2005by}. Let $d(n)$ represent the number of configurations of the heterotic string with electric charge such that $ \tfrac12 \mathbf{q}_e^2=n$. The level matching condition, $n= \tfrac12 \mathbf{q}_e^2=N_L-1$, implies that we need to count the number of states with total oscillator number $N_L=(n+1)$. The generating function for such states is
\begin{equation}
\frac{16}{\eta(\tau)^{24}} = \sum_{n=-1}^\infty d(n) \ q^n\ ,
\end{equation}
where the factor of $16$ accounts for the degeneracy of a $\tfrac12$-BPS multiplet -- this is the degeneracy of the Ramond ground state in the right-moving sector. For the orbifolded models, this partition function gets modified because of the presence of twisted sectors and one needs to add the contribution from the different sectors to get the correct partition function. Sen has studied the degeneracy of the $\tfrac12$-BPS states in the orbifolded models and showed that, up to exponentially suppressed terms (for large charges), the leading contribution arises from the twisted sectors. It turns out, for the $\mathbb Z_N$-orbifolded theories with $N$ being prime and subject to the constraint $(k+2) = 24/(N+1)$, the generating functions for the $\tfrac12$-BPS states are just the unique cusp forms for $\Gamma_0(N)$. In \cite{Govindarajan:2009qt} extending and generalizing Sen's result, a more general ansatz for the generating function for the $\frac12$-BPS states was given based on the relation of the symplectic automorphisms of $K3$ with the conjugacy classes of $M_{24}$. The generating functions for the degeneracy of the $\tfrac12$-BPS states in the orbifolded models, for both prime and composite values of $N$, were found to be given by multiplicative $\eta$-products, of weight $k+2$, associated with specific (balanced) cycle shapes corresponding to the conjugacy classes of the $24$-dimensional permutation representation of $M_{24}$. A detailed discussion of the cycle shapes and $\eta$-products leading to the generating functions of half-BPS states can be found in \cite{Govindarajan:2009qt, Govindarajan:2010ge}. Taking into account the fact that the electric charge is quantized such that $N \mathbf{q}_e^2\in 2\mathbb{Z}$ in the $\mathbb Z_N$-orbifolded theories, one finds that the degeneracies of the $\tfrac12$-BPS states are generated by the $\eta$-products as
\begin{equation}
\frac{16}{g_\rho(\tau/N)}\equiv \sum_{n=-1}^\infty d(n) \ q^{n/N}\ ,
\end{equation}
for the $\mathbb{Z}_N$ CHL orbifold. The subscript $\rho$ corresponds to the $M_{24}$ cycle shape from which the multiplicative $\eta$-product $g_\rho(\tau)$ is constructed. The generating functions $g_\rho(\tau)$ are cusp forms of $\Gamma_0(N)$ for the $\mathbb Z_N$-orbifolded theory just like $\eta(\tau)^{24}$ is a cusp form of the modular group $\Gamma$. This is not unexpected because, for the orbifolded theories, the S-duality group is no longer $SL(2, \mathbb Z)$ but is broken down to a smaller subgroup, $\Gamma_1(N)$. Correspondingly, the cusp forms generating the $\tfrac12$-BPS degeneracies are broken down from the cusp form of $\Gamma$ to cusp forms of its subgroups. 

The degeneracy of the $\tfrac12$-BPS states in the $N=5$ case are generated by the weight $4$ cusp form $1/ \eta(\tau)^4 \eta(\tau/5)^4$. The degeneracy is given by 
\begin{equation}
\frac{16}{\eta(\tau)^4 \eta(\tau/5)^4} = \sum_{n=-1}^\infty d(n) \ q^{n/5}\ .
\end{equation}
Analogous to the DVV case, this $\eta$-product is an input into the genus-two Siegel modular that generates the $\tfrac14$-BPS degeneracies. It forms a part of the seed for the additive lift generating the genus-two Siegel modular form as an infinite sum via its Fourier-Jacobi expansion.

The product of $g_\rho(\tau)$ with $\tfrac{\vartheta_1(z_1,z_2)^2}{\eta(z_1)^6}$ gives a weak Jacobi form of weight $k$, index $1$ and level $N$
\begin{equation}
\label{additiveseed}
\phi_{k,1}(z_1,z_2) = \frac{\vartheta_1(z_1,z_2)^2}{\eta(z_1)^6} \ g_\rho(z_1)=
\sum_{n,\ell} a(n,\ell)\ q^n r^\ell ,
\end{equation}
which is the additive seed for generating the modular form $\Phi_2 (\mathbf Z)$. The modular form generating the $\tfrac14$-BPS degeneracies, $\widetilde{\Phi}_k(\mathbf{Z})$, is given by expanding the modular form $ \Phi_k(\mathbf{Z})$ about another inequivalent cusp. The modular form $\widetilde{\Phi}_k(\mathbf{Z})$ is related to the modular form $\Phi_2 (\mathbf Z)$ as 
\begin{equation}
\widetilde{\Phi}_k(\mathbf Z) = \ z_1^{-k} \
\Phi_k(\mathbf{\widetilde{Z}})\ ,
\end{equation}
with
\[
\tilde{z}_1 = -1/z_1\quad,\quad \tilde{z}_2 = z_2/z_1\quad, \quad 
\tilde{z}_3 = z_3 -z_2^2/z_1\ .
\]
The additive seed for $\widetilde{\Phi}_k(\mathbf{Z})$ is thus given by the weak Jacobi form $\phi_{k, 1}(\tfrac{-1}{z_1}, \tfrac{z_2}{z_1})$ and the genus-two Siegel modular form is generated from it after summing over the index $m$ for all values of $m \geq 1$ as\cite{Jatkar:2005bh} 
\begin{equation}
\widetilde{\Phi}_k (z_1, z_2, z_3) = \sum_{m \geq 1} e^{2\pi i m z_3} z_1^{-k} e^{-2 \pi i m z_2^2 /z_1} \phi_{k,m}(\tfrac{-1}{z_1}, \tfrac{z_2}{z_1}) \ .
\end{equation}

\noindent For the $\mathbb Z_5$-orbifold, the additive seed is given by 
\begin{equation}
\phi_{2,1}(\tfrac{-1}{z_1}, \tfrac{z_2}{z_1}) = \frac{\vartheta_1(z_1,z_2)^2}{\eta(z_1)^6} \times g_\rho(z_1/N)\eta(\tau)^4 \eta(\tau/5)^4 \ .
\end{equation}
Following the procedure of Jatkar and Sen\cite{Jatkar:2005bh}, the genus-two Siegel modular form $\widetilde{\Phi}_2 (\mathbf Z)$ would then be obtained as 
\begin{equation}
\widetilde{\Phi}_2 (z_1, z_2, z_3) = \sum_{m \geq 1} e^{2\pi i m z_3} \ \frac{\vartheta_1(z_1,z_2)^2}{\eta(z_1)^2} \ \eta(\tau/5)^4 \ .
\end{equation}

\noindent However, the seed for the $N=5$ case is a weak Jacobi form, and the validity of the above procedure is not guaranteed. One needs to verify the expansion of the modular form independently from a different procedure.The generalized theta correspondence gives another method to obtain the same modular form, this time as an infinite product. This is useful not just as a check for the modular form constructed via the additive lift, but also in interpreting the modular form as the Weyl-Kac-Borcherds denominator formula of the BKM Lie superalgebra. Now we discuss the product representation of the modular form $\widetilde{\Phi}_k(\mathbf{Z})$.


\subsection{Product formulae}

Product representations for the genus-two Siegel modular forms, $\Phi_k (\mathbb Z)$ and $\widetilde{\Phi}_k (\mathbb Z)$, can be obtained from string threshold correction computations \cite{Dixon:1990pc, Kawai:1993jk, Kawai:1995hy, Harvey:1996gc, Harvey:1995fq} and for the modular forms occurring in the CHL theories was computed by David, Jatkar and Sen\cite{David:2006ji} using essentially the same method as \cite{Kawai:1995hy}. Upon evaluating the integral and requiring its invariance under the duality transformations one obtains the modular forms $\Phi_k(\mathbf{Z})$ and $\widetilde{\Phi}_k(\mathbf{Z})$ as an infinite product given in terms of the coefficients of the Fourier expansion of the twisted elliptic genera of $K3$. 

 This essentially is the generalized theta correspondence of Borcherds specialized to the case of lattices $\mathbb R^{2, s}$ giving holomorphic automorphic forms as infinite products on the hermitian symmetric space $G(2, s)$ (Theorem $13.3$, \cite{Borcherds:1998Aut}). The correspondence relates a holomorphic (vector valued) modular form to automorphic forms on $O_{2, s}$ given as an infinite product with the coefficients coming from the vector valued modular form. In this case the log of the automorphic form is obtained from the generalized theta correspondence and the weight of the automorphic form is given by the zeroth coefficient in the Fourier expansion of the vector valued modular form. In the present case the vector valued modular form is the twisted elliptic genus for $K3$ orbifolded by a $ \mathbb{Z}_N $ group. 

\noindent The twisted elliptic genus for a $ \mathbb{Z}_N $-orbifold of $K3$ is defined as:
\begin{equation}
F^{a,b}(\tau,z) = \frac1N \textrm{Tr}_{RR,g^a} \Big((-)^{F_L+F_R} g^b q^{L_0}\bar{q}^{\bar{L}_0} e^{2\pi\imath z F_L}\Big)\ ,\quad 0\leq a\leq (N-1)\ ,
\end{equation}
where $g$ generates the $\mathbb{Z}_N$ transformation and $ q=\exp(2\pi i \tau)$. These are weak Jacobi forms of weight zero, index one and level $N$\cite{David:2006ji}. A weak Jacobi form, $F^{0,s}(\tau,z)$, of $\Gamma_0^J (N)$\footnote{The Jacobi group, $\Gamma^J = SL(2, \mathbb Z) \ltimes H(\mathbb Z)$, is the sub-group of Sp$(2, \mathbb Z)$ that preserves the cusp at $z_3 = i \infty$. $H(\mathbb Z)$ is the Heisenberg group. The group $\Gamma_0^J (N)$ corresponds to the congruence subgroup obtained by considering the congruence subgroup $\Gamma_0(N)$ in place of $SL(2, \mathbb Z)$.}, can be written as\cite{Aoki:2005}
\begin{eqnarray}
\label{Frsgeneral}
F^{0,0}(\tau,z)&=& \tfrac2N A(\tau,z)\quad, \\
F^{0,s}(\tau,z) &=& a_s\ A(\tau,z) + \alpha_{N,s}(\tau)\ B(\tau,z)\ ,\ \ s\neq 0\ ,
\end{eqnarray}
where $ \alpha_{N,s}(\tau) $ is a weight-two modular form of $\Gamma_0(N)$ and
\begin{equation}\label{ABdef}
A(z_1,z_2)
=  \sum_{i=2}^4 \left(\frac{\vartheta_i(z_1,z_2)}{\vartheta_i(z_1,0)}\right)^2 
\quad,\quad
B(z_1,z_2)=
\left(\frac{\vartheta_1(z_1,z_2)}{\eta^3(z_1)}\right)^2 \ .
\end{equation}
For prime $N$, the modular forms $\alpha_{N,s}(\tau)$ at weight two is generated by the weight two holomorphic Eisenstein series of $\Gamma_0(N)$, $E_4(\tau)$, defined from the weight two non-holomorphic modular form of $SL(2, \mathbb Z)$ as 
\begin{equation}
E_{N}(\tau)=\frac1{N-1}\Big(N E_2^*(N\tau)-E_2^*(\tau)\Big)= \tfrac{12i}{\pi(N-1)}\partial_\tau \big[\ln 
\eta(\tau) -\ln \eta(N\tau)\big],
\end{equation}
with constant coefficient equal to $1$\cite[Theorem 5.8]{Stein}. The subscript $N$ indicates the level and not the weight of the Eisenstein series, which is two. Also, the action of an $SL(2, \mathbb Z)$ element on the weight $0$ modular forms, $F^{a,b}(\tau,z)$, is given by
\begin{equation}
\label{Frstransformation}
F^{a,b}\Big(\frac{a\tau +b}{c\tau +d},\frac{z}{c\tau +d}\Big) = \textrm{exp}\Big(2\pi i \frac{cz^2}{c\tau +d}\Big) F^{cs+ar,ds+br}(\tau,z), 
\end{equation}
for $a, b, c, d \in \mathbb Z$ and $ad-bc =1$. Thus, for prime $N$, using \eqref{Frsgeneral} and \eqref{Frstransformation} one can get the Fourier coefficients of the weight zero weak Jacobi forms, $F(r, s)$ for all $r, s$, knowing the weight two Eisenstein series, $E_N (\tau)$, at level $N$. The Fourier expansion of the Jacobi forms are 
\begin{equation}
F^{a,b}(\tau,z)= \sum_{m=0}^1 \sum_{\substack{\ell\in 2\mathbb{Z}+m,\\ n\in \mathbb{Z}/N}}c_m^{a,b}(4n-\ell^2)\ q^n r^\ell\ , \label{fourierdefs}
\end{equation}
where  $r=\exp(2\pi i z)$. Using the Fourier coefficients, the product formula for the modular form, $\widetilde{\Phi}_2 (\mathbf Z)$, generating the degeneracies of $\tfrac14$-BPS states in the $\mathbb Z_5$-orbifolded model is given by
\begin{align}
 \widetilde{\Phi}_2(\mathbf{Z}) = q^{1/5} r s  \prod_{a=0}^{4} \prod_{\substack{\ell,m\in \BZ,\\ n \in \BZ \pm
\tfrac{a}{5}}} \Big(1- q^nr^\ell s^m\Big)^{\sum_{b=0}^4 \omega^{\mp bm}
c^{(a,b)}(4nm -\ell^2)}
\end{align}
where $\omega=\exp(\tfrac{2\pi\imath}4)$ is a fifth-root of unity, and $c^{(a,b)}(4nm- \ell^2)$ are the Fourier coefficients of the twisted elliptic genera, $F^{(a,b)}(z_1,z_2)$ given in \eqref{fourierdefs}. 

\subsection{Comparing the additive and multiplicative lifts}

Constructing the modular form via two different methods to generate it as an infinite sum and an infinite product affords a non-trivial check for the validity of the methods. It is also necessary to construct the BKM Lie superalgebra from the modular form via the denominator formula.  Comparing the modular forms generated from the additive and multiplicative lifts, we find a minor mismatch between the two expansions. The additive and multiplicative expansions do not match with each other. In such a case, one needs another way of verifying which of the two expansions is the one generating the dyon degeneracies.

The correspondence between the roots of the BKM Lie superalgebra, constructed from the square roots of the modular forms, and the walls of marginal stability of the $\tfrac14$-BPS states provides another minor check for the modular forms constructed from the additive and multiplicative lifts. Under the correspondence, the BKM Lie superalgebra constructed from the modular form should have a one-to-one correspondence between its real simple roots and the walls of marginal stability of the $\tfrac14$-BPS states for the model. Comparing with the analysis coming from the correspondence with the walls of marginal stability suggests that the product formula gives the correct modular form and one needs to add terms to the modular form generated from the additive lift, to make it match the one generated from the multiplicative lift and give the correct correspondence with the walls of marginal stability. The case of the $N=5$ CHL theory is different from the $N=1, 2, 3, 4$ because the seed for the additive lift in this case is a weak Jacobi form, where for $N<5$ it was a Jacobi form. 

The walls of marginal stability for the $\mathbb Z_5$-orbifolded theory are infinite in number, and divided into two chambers separated by two limit points. The appearance of a second chamber is not seen in the $N \leq 4$ theories, and is seen for the first time in the $N=5$ theory. Taking the correspondence on the BKM Lie superalgebra side, one finds that the two chambers generate two sets of equivalent roots which are in one-to-one correspondence with each other. While the modular form constructed from the product formula contains all the roots coming from both the chambers, the modular form constructed from the additive lift does not seem to contain the roots arising from the new chamber, while containing all the roots from the old chamber. We now discuss the above ideas in detail, starting with the BKM Lie superalgebras. 

\section{BKM Lie superalgebra}

BKM (Borcherds-Kac-Moody) Lie superalgebras are the most general class of (infinite dimensional)-Lie algebras\cite{Borcherds:1990, Ray:2006}. They were first constructed by Borcherds in the context of Conway and Norton's moonshine conjecture for the Monster group, extending the theory of Kac-Moody algebras. In the present situation, BKM Lie superalgebras appear in relation to the modular forms (more precisely, their square roots) constructed in the previous section via their Weyl-Kac-Borcherds denominator identities. 

It was already known that the genus-two Siegel modular form, $\Phi_{10}(\mathbf Z)$, constructed in the context of generating dyonic degeneracies\cite{Dijkgraaf:1996it} or from the threshold corrections\cite{Kawai:1995hy} was related to a BKM Lie superalgebra constructed by Gritsenko and Nikulin\cite{Nikulin:1995}. So, when the modular forms generating the degeneracy of dyonic states in the CHL models were constructed, it was natural to look for a possible BKM Lie superalgebra structure related to them. Progress in that direction was carried out in\cite{Govindarajan:2008vi, Govindarajan:2009qt, Cheng:2008kt, Govindarajan:2010ge} where the two families of rank $3$ BKM Lie superalgebras, that correspond to the two modular forms $\widetilde{\Phi}_k (\mathbf Z)$ and $\Phi_k (\mathbf Z)$ for the N$=2,3,4$ CHL theories were shown to exist. 

The modular forms are related to the BKM Lie superalgebras via their denominator formulae. The  Weyl-Kac-Borcherds (WKB) denominator formula is the special case of the more general WKB character formula for Lie algebras which gives the characters of integrable highest weight representations of BKM Lie superalgebras. The WKB character formula applied to the trivial representation gives the WKB denominator formula. It contains all the information of the algebra, and knowing the denominator formula one can reconstruct the algebra from it. It is given as an equality between an infinite sum and an infinite product constructed out of the roots, their multiplicities, and the Weyl group action. The two sides of the equality are equated to the sum and product representations of the modular forms constructed in the previous sections. 

Let $\mathfrak{g}$ be a BKM Lie superalgebra and $\mathcal{W}$ its Weyl group. Let $L_+$ denote the set of positive roots of the BKM Lie superalgebra and $\rho$ the Weyl vector. Then, the WKB denominator identity for the BKM Lie superalgebra $\mathfrak{g}$ is
\begin{equation}
\label{WKB}
\prod_{\alpha \in L_+} (1- e^{- \alpha})^{\textrm{mult}(\alpha)} = e^{-\rho}\
\sum_{w \in \mathcal{W}} (\det w)\  w(e^{\rho} \sum_{\alpha \in L_+}
\epsilon (\alpha) e^{\alpha} )\  , 
\end{equation}
where  mult$(\alpha)$ is the multiplicity of a root $\alpha \in L_+$. In the above equation, det($w$) is defined to be $\pm1$ depending on whether $w$ is the product of an even or odd number of reflections and $\epsilon (\alpha)$ is defined to be $(-1)^n$ if $\alpha$ is the sum of $n$ pairwise independent, orthogonal imaginary simple roots, and $0$ otherwise. In the case of BKM Lie superalgebras the roots appear with graded multiplicity -- fermionic roots appear with negative multiplicity while bosonic roots appear with positive multiplicity. 

The BKM Lie superalgebras for the $N=1,2, 3$ CHL models were all elliptic, with $3$, $4$ and $6$ real simple roots respectively. They are given by the Cartan matrices 
\begin{equation}
\label{CartanN1}
A_{1,II}=\begin{pmatrix}
2 & -2 & -2 \\
-2 & 2 & -2 \\
-2 & -2 & 2
\end{pmatrix}\ .
\end{equation}
\begin{equation}
\label{CartanN2}
A_{2,II}=\begin{pmatrix}
2 & -2 & -6 & -2 \\
-2 & 2 & -2 & -6 \\
-6 &-2 & 2 & -2  \\
-2 & -6 & -2 & 2 
\end{pmatrix}\ .
\end{equation}
\begin{equation}
\label{CartanN3}
A_{3,II}=\begin{pmatrix}
2 & -2 & -10 & -14 & -10 & -2 \\
-2 & 2 & -2 & -10 & -14 & -10 \\
-10 & -2 & 2 & -2 & -10 & -14 \\
-14 & -10 & -2 & 2 & -2 & -10 \\
-10 & -14 & -10 & -2 & 2 & -2 \\
-2 & -10 & -14 & -10 & -2 & 2
\end{pmatrix}\ .
\end{equation}

The BKM Lie superalgebra for the $N=4$ model is of parabolic nature with an infinite number of real simple roots in the algebra. This was in agreement with the fact that the CHL theory with $\mathbb Z_4$-obrifolding was known to have an infinite number of walls of marginal stability\cite{Sen:2007vb} and there exists a one-to-one correspondence between the walls of marginal stability and the real simple roots of the corresponding BKM algebra. The Cartan matrix of the BKM Lie superalgebra $\widetilde{\mathcal G}_4$ is given by 
\begin{equation}\label{CartanN4}
A^{(4)} = (a_{nm})\quad \textrm{where}\quad 
a_{nm}= 2 -4(n-m)^2\  ,
\end{equation}
with $m,n\in \mathbb{Z}$.

For the $\mathbb Z_5$-obrifold, one finds that there exists a BKM Lie superalgebra, again with an infinite number of real simple roots, but of hyperbolic nature. This is the first instance of a hyperbolic BKM Lie superalgebra. It is given by the Cartan matrix (see section \ref{WallWeylChamber}, eq. \eqref{infiniteCartan})
\begin{equation}
A^{(5)} = (a_{nm})\quad \textrm{where}\quad 
a_{nm}= - 4 \Big( \Lambda_1^{-n} \Lambda_2^{-m} + \Lambda_1^{-m} \Lambda_2^{-n} \Big) + (-1)^{d(x_m) + d(x_n)} 10, \nonumber
\end{equation}
where $m,n\in \mathbb{Z}$, $\Lambda_1$ and $\Lambda_2$ are the roots of the equation $r^2 - 3 \cdot r + 1 = 0$ and $d(x_m) = 0$ if $x_m \in \mathcal B_L \cup \mathcal B_R$ and $d(x_m) = 1$ if $x_m \in \mathcal B_C$ is the grading on the two sets of roots of the algebra, and $ \mathcal B_L \cup \mathcal B_R$ and $ \mathcal B_C$ are the two chambers in the fundamental domain of the walls of marginal stability. We will come back to discuss the algebra $\mathcal{\widetilde G}_5$ in more detail again after studying the walls of marginal stability for the $\tfrac14$-BPS states in the model. We will see a detailed construction of the above given Cartan matrix and also point out certain peculiar aspects of the algebra in relation to the ones constructed before it.  

\section{Walls of marginal stability}

Now we come to an important idea which establishes a relation between the algebras discussed in the previous section and the string theories they are constructed from. Such a bridge is important because our final aim is to understand the origin of these algebras and the role played by them in the theory. Hence one needs to construct a dictionary between the algebraic and physical sides. One such relation was found by Cheng and Verlinde\cite{Cheng:2008fc} between the walls of marginal stability for the $\tfrac14$-states and the walls of the Weyl chambers of the algebras. In this section we study this in the case of the $\mathbb Z_5$-orbifold before studying the more general case. 

The walls of marginal stability for the $\tfrac14$-BPS states in $\mathcal N=4$ supersymmetric theories was originally studied by Sen\cite{Sen:2007vb}(see also\cite{Mukherjee:2007nc,Mukherjee:2007af}). The CHL theory for the $\mathbb Z_5$-orbifold has an infinite number of walls of marginal stability just like the $N=4$ case. We will eventually see that this is part of a pattern that exists across the CHL theories with orbifolding groups $\mathbb Z_N$ for $N=4, 5, 6$ and $7$, all of which have an infinite number of walls of marginal stability. Before taking a bird's eye view of the walls of marginal stability across different $N$, however, we will first start with a detailed study of the $N=5$ model.

For dyons with a given set of charges, the moduli space of $\mathcal N=4$ supersymmetric theories contains subspaces -- of co-dimension one, known as the walls of marginal stability -- on which the dyon becomes marginally unstable against decay into two $\tfrac12$-BPS states respecting the conservation of charge and mass across the subspace. The spectrum of the $\tfrac14$-BPS states therefore changes discontinuously as one moves through a wall of marginal stability in the moduli space. The degeneracy formula is thus valid only in a limited domain in the moduli space. As one crosses a wall of marginal stability, one needs to take into account the fact that a $\tfrac14$-BPS state may split into two $\tfrac12$-BPS states under suitable circumstances. However, the change in the expression for the degeneracy across the wall is not very drastic and is such that the partition function for the $\tfrac14$-BPS degeneracies formally remains the same, but the point in the Siegel moduli space around which one should series expand the partition function to extract the degeneracies changes as one moves across a wall. From the point of view of the BKM Lie superalgebras, it is a change in the Weyl chamber of the algebra, and the set of real simple roots. 
 
The walls are, in general, given by complicated dependence on all the moduli in the moduli space. However, for the sake of simplicity, one can study them in the axion-dilaton moduli space fixing all the other moduli. The walls of marginal stability would then be curves in the axion-dilaton plane (which is modelled by the complex upper half-plane). The curves for the walls of marginal stability are found to be circles and straight lines in the upper half-plane which intersect only on the real $\lambda$-axis or at $i \infty$, but no where in the interior of the upper half-plane. The points of intersection have a universal nature, in that, although the qualitative features of the curves depend on the charges and other moduli, the points of intersection on the real $\lambda$-axis depend only on $N$. 

\begin{figure}[h]
\centering
\includegraphics[height=2in]{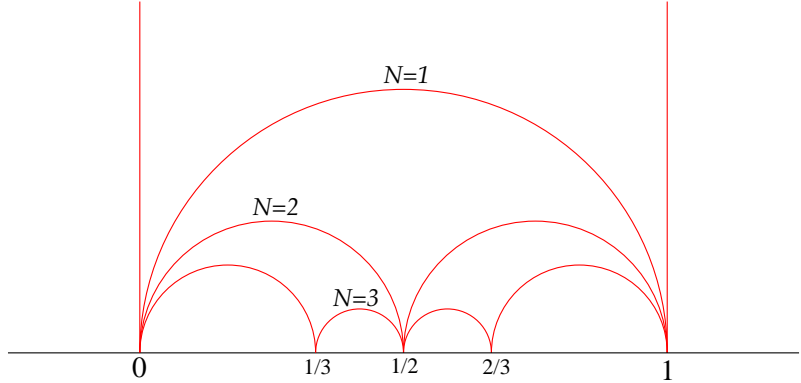}
\caption{Fundamental domains/Weyl chambers for $N=1,2,3$}
\label{weylchamber123}
\end{figure}

One can study the walls of marginal stability in a fundamental domain in the upper half plane by restricting the value of Re$(\lambda)$ to the interval $[0,1]$. The straight lines Re$(\lambda)=0,1$ correspond to two walls. The other walls are semi-circles intersecting each other on the Re$(\lambda)$-axis between $[0,1]$. For the case of $N=1,2,3$ one has only a finite number of walls in the $[0,1]$ interval with the intersection points given by the following set of intercepts on the real line:
\begin{equation}
(\tfrac01,\tfrac11)\ ,\quad (\tfrac01,\tfrac12,\tfrac11)\ , \quad (\tfrac01,\tfrac13, \tfrac12,\tfrac23,\tfrac11)\ .
\end{equation}
A fundamental domain is then given by restricting to the region bounded by these semi-circles and the two walls connecting $\lambda=0,1$ to infinity. The two straight lines may be included by adding the `points' $\tfrac{-1}0$ and $\tfrac10$. 
The fundamental domains for the $N=1,2,3$ models are given in Figure \ref{weylchamber123}.

\begin{figure}[h]
\centering
\includegraphics[height=2in]{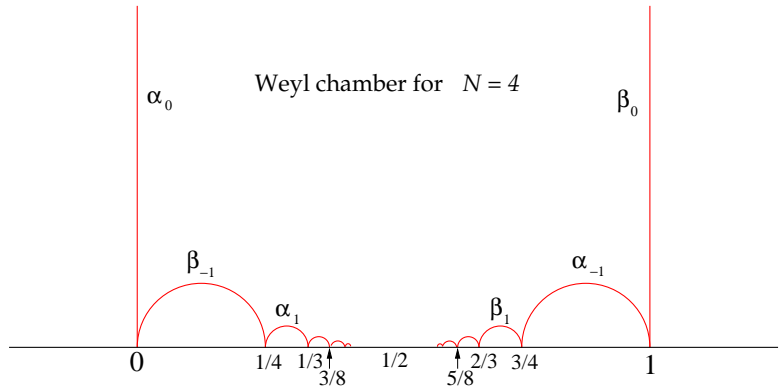}
\caption{Fundamental domains/Weyl chambers for $N=4$}
\label{weylchamber4}
\end{figure}

For $N>3$, this picture does not terminate -- one needs an infinite number of semi-circles to obtain a closed domain. The walls of marginal stability for the $\mathbb Z_4$-orbifold was studied in\cite{Govindarajan:2009qt, Sen:2007vb} where it was found that the intercepts of the walls on the Re($\lambda$)-axis are given by 
\begin{equation}
\label{Nequals4walls}
(\tfrac01,\tfrac14,\tfrac13,\tfrac38,\tfrac25,\ldots,\tfrac{-2n+1}{-4n},\tfrac{-n}{-2n-1},\ldots,\tfrac12,\ldots,\tfrac{n+1}{2n+1},\tfrac{2n+1}{4n}\ldots,\tfrac35,\tfrac58,\tfrac23,\tfrac34,\tfrac11)\ .
\end{equation}
There are an infinite number of intercepts corresponding to an infinite number of walls in the fundamental domain. This agrees with the fact that the corresponding BKM Lie superalgebra has an infinite number of real simple roots. The fundamental domain for $N=4$ is given in Figure \ref{weylchamber4}. 

\subsection{Walls of marginal stability for the $\mathbb Z_5$-orbifold}

For the $N=5$ case, following Sen's method\cite{Sen:2007vb} to find the location of the walls, one again obtains an infinite number of walls in the fundamental domain. However, the fundamental domain is divided into three regions, denoted $\mathcal B_R, \mathcal B_L$, and $\mathcal B_C$, where for the cases $N = 2, 3, 4$ there were only two domains, namely $\mathcal B_R$ and $\mathcal B_L$ which were symmetric about the point $\tfrac12$, and for $N=1$, only one domain. The appearance of a new region, $\mathcal B_C$, happens only for $N \geq 5$. For the $N= 2, 3, 4$ cases the point $\tfrac12$ was a limit point for the set of walls starting at $0$ and $1$. For $N=5$, one has two limit points given by $\tfrac12 (1 \pm \sqrt{\tfrac15})$ corresponding to the three chambers that the fundamental domain is divided into. 

The intercepts of the walls, on the Re($\lambda$)-axis in the three domains are given by
\begin{equation}\label{lchamber5}
\mathcal B_L = \big \{ \frac{A_n}{B_n} , \frac{B_n}{5 \cdot A_{n+1}} \big \} , n \in \mathbb{Z}_+ ,
\end{equation}
starting at the intercept $\tfrac01$,
\begin{equation}\label{rchamber5}
\mathcal B_R = \big \{ \frac{A_{n+1}}{B_n} , \frac{B_{n+1}}{5 \cdot A_{n+1}} \big \} , n \in \mathbb{Z}_+ ,
\end{equation}
starting at the intercept\footnote{In fact, the two chambers $\mathcal B_L$ and $\mathcal B_R$ can be considered as one chamber by allowing $n \in \mathbb Z$ instead of $\mathbb{Z}_+$, and from here on we think of $\mathcal B_L \cup \mathcal B_R$ as one connected chamber.} $\tfrac11$, where,
\begin{equation}
A_n = (\tfrac1{\sqrt 5}) \Lambda_1^n - (\tfrac1{\sqrt 5}) \Lambda_2^n, \quad B_n = (\tfrac{1+ \sqrt 5}2) \Lambda_1^n + (\tfrac{1- \sqrt 5}2) \Lambda_2^n  ,
\end{equation}
while the intercepts of the walls on the Re($\lambda$)-axis in the domain $\mathcal B_C$ are given by 
\begin{equation}
\mathcal B_C = \big \{ \frac{\tilde{A}_n}{\tilde{B}_n} , \frac{\tilde{B}_{n+1}}{5 \cdot \tilde{A}_n} \big \} ,
\end{equation}
where, 
\begin{equation}
\label{tildeAB}
\tilde{A}_n = (\tfrac{\sqrt 5-1}{2 \sqrt 5}) \Lambda_1^n + (\tfrac{\sqrt 5+1}{2 \sqrt 5}) \Lambda_2^n, \quad \tilde{B}_n = (\tfrac{3- \sqrt 5}2) \Lambda_1^n + (\tfrac{3+ \sqrt 5}2) \Lambda_2^n,
\end{equation}
and $\Lambda_1$, $\Lambda_2$ in all the equations above are given by
\begin{equation}
\label{rootsofchambers}
\Lambda_1 = \tfrac{3+ \sqrt 5}2, \quad \Lambda_2 = \tfrac{3- \sqrt 5}2 \ .
\end{equation}
The limit points of these ratios of sequences are as expected by Sen in his analysis of the walls of marginal stability\cite{Sen:2007vb}. 

Since there are an infinite number of walls, having a systematic way of labeling helps in studying them. The labeling is also very important to write down the Cartan matrix of the associated BKM Lie superalgebra. It is clear from the structure of the walls that we will need two separate sequences to label the walls. Following \cite{Govindarajan:2009qt} we label the walls by two sequences, $\alpha_n$ and $\beta_n$, indexed by an integer, $n$. The sequences, $\alpha_n$ and $\beta_n$, denote the semi-circles in $\mathcal B_L \cup \mathcal B_R$ with intercepts given by 
\begin{equation}
\alpha_n = \begin{pmatrix} B_{n-1} & A_n \\ 5 \cdot A_{n} & B_{n} \end{pmatrix} \ \textrm{ and } \ \beta_n = \begin{pmatrix}  A_{n+1} & B_{n} \\ B_{n} & 5 \cdot A_{n} \end{pmatrix} \ .
\end{equation}
It will turn out that these correspond to the real simple roots of the BKM Lie superalgebra $\widetilde{\mathcal G}_5$. Note that $\alpha_0$ and $\beta_0$ represent the two straight lines at Re$(\lambda)=0,1$ respectively. 

Like one did in the $N=4$ case, one can interpret the fundamental domain as a regular polygon with an infinite number of edges and with an infinite-dimensional dihedral group, $D_\infty$, as its symmetry group. In this case, there are two polygons corresponding to the two sets of chambers $\mathcal B_L \cup \mathcal B_R$ and $\mathcal B_C$. Each of the polygons has two infinite-dimensional dihedral symmetry groups denoted $D^{(1)}_\infty$ and $D^{(2)}_\infty$. The group $D^{(1)}_\infty$ is generated by two generators: a reflection $y$ and a shift $\gamma$ given by:
\begin{equation}
\label{Dinfinity}
y:\quad  \alpha_n \rightarrow \alpha_{-n}\ ,\ \beta_n \rightarrow \beta_{-n-1}\quad \textrm{and}\quad  \gamma: \alpha_n \rightarrow \alpha_{n+1}\ ,\ 
\beta_n \rightarrow \beta_{n-1}\ ,
\end{equation}
satisfying the relations $y^2=1$ and $y\cdot \gamma\cdot y=\gamma^{-1}$. The transformation $\gamma \in \widetilde{\Gamma}_1 (N)$\footnote{The extended S-duality group, $\widetilde{\Gamma}_1 (N)$, is defined by including a $\mathbb{Z}_2$ parity operation to the S-duality group $\Gamma_1(N)$. For $N=1$, this is the group $PGL(2,\mathbb{Z})$\cite{Cheng:2008fc}. The generator $y$ is not realized as an element of a level $4$ subgroup of $PGL(2,\BZ)$ and thus is not an element of the extended S-duality group. This is similar to what happens for $N=2,3$ and $4$.} permutes the set of walls in $\mathcal B_L \cup \mathcal B_R$, and hence the real simple roots of the BKM Lie superalgebra, just like for the $N<5$ cases. Its action on the walls in the central chamber, $\mathcal B_C$, is the same as given in eq. \eqref{Dinfinity} for the walls in the chamber $\mathcal B_L \cup \mathcal B_R$. The transformation $\gamma$ is realised as a $\Gamma_0(5)$ matrix as $\gamma = \begin{pmatrix} 1 & -1 \\ N & 1-N \end{pmatrix} = \begin{pmatrix} 1 & -1 \\ 5 & -4 \end{pmatrix}$\cite{Cheng:2008kt}.

The transformation, $\delta$, that exchanges the walls $\alpha_n$ and $\beta_n$ generates a second $\BZ_2$ defined as follows:
\begin{equation}
\label{deltaaction}
\delta= \begin{pmatrix} -1 & 1 \\ 0 &1 \end{pmatrix}:\quad \alpha_n \longleftrightarrow \beta_n\ .
\end{equation}
The transformations $(\gamma,\delta)$ together generate the other dihedral group denoted $D^{(2)}_\infty$. 

Similarly, choosing a labeling for the roots occurring in the chamber $\mathcal B_C$, the walls can be written as two infinite sequences, $\tilde{\alpha}_n$ and $\tilde{\beta}_n$, labelled by an integer $n$ as
\begin{equation}
\tilde{\alpha}_n = \begin{pmatrix} \tilde{B}_{-n+1} & \tilde{A}_{-n+1} \\ 5 \cdot \tilde{A}_{-n} & \tilde{B}_{-n+1} \end{pmatrix} \ \textrm{ and } \ \tilde{\beta}_n = \begin{pmatrix}  \tilde{A}_{n+1} & \tilde{B}_{n+2} \\ \tilde{B}_{n+1} & 5 \cdot \tilde{A}_{n+1} \end{pmatrix},
\end{equation}
where $\tilde{A}_n$ and $\tilde{B}_n$ are as given in eq. \eqref{tildeAB}. The chamber $\mathcal B_C$ also has a dihedral symmetry group given by $D^{(1)}_\infty$ as defined in \eqref{Dinfinity}. The second dihedral group, $D^{(2)}_\infty$, generated by $\delta$ and $\gamma$ is also a symmetry of the chamber $\mathcal B_C$, with its action on the roots as given in \eqref{deltaaction} with $\alpha_n, \beta_n$ replaced by $\widetilde{\alpha}_n$ and $\widetilde{\beta}_n$. Thus, there would appear to be two seperate chambers given by $\mathcal B_L \cup \mathcal B_R$ and $\mathcal B_C$, each an infinite polygon with the same set of dihedral symmetry groups. 

This can be understood as follows. Although the intercepts on the real $\lambda$-axis in the two sets $\mathcal B_L \cup \mathcal B_R$ and $\mathcal B_C$ look completely different, following the correspondence on the BKM Lie superalgebra side one can see that the two polygons are indeed identical. In the BKM Lie superalgebra, $\widetilde{\mathcal G}_5$, the sets of roots arising from the two polygons are in exact one-to-one correspondence with each other. With the labeling introduced earlier, for each root $\alpha_n$ in the polygon $\mathcal B_L \cup \mathcal B_R$, there exists a corresponding root $\widetilde{\alpha}_n$ in the polygon $\mathcal B_C$ and similarly with the roots $\beta_n$ and $\widetilde{\beta}_n$. Further, computing the element in the Cartan matrix corresponding to the inner product between any two roots in each of the two chambers, one finds that
\begin{equation}
(x_n, x_m) = (\widetilde{x}_n, \widetilde{x}_m), \ x_i \in \mathcal B_L \cup \mathcal B_R, \ \widetilde{x}_i \in \mathcal B_C \ .
\end{equation}
That is, the inner product between any two roots in the polygon $\mathcal B_L \cup \mathcal B_R$ is the same as that for the corresponding two roots in the polygon $\mathcal B_C$. Thus, one sees that the two polygons are identical to each other and give rise to two identical copies of the roots in the BKM Lie superalgebra with the same inner products. Hence it is intuitively understandable that they are same as infinite polygons with the same sets of dihedral symmetry groups. 

One would also need to find transformations that take the walls from $\mathcal B_L \cup \mathcal B_R$ chamber to the chamber $\mathcal B_C$. This is given by the $\Gamma_0(5)$ matrix, $\sigma$, with determinant $+1$ as follows\cite{Sen:2007vb}
\begin{equation}
\sigma  = \begin{pmatrix} 2 & -1 \\ 5& -2 \end{pmatrix}: \alpha_n \leftrightarrow \tilde{\alpha}_{-n}, \quad \beta_n \leftrightarrow \tilde{\beta}_{-n-1}, \textrm{ where } \alpha_n, \beta_n \in \mathcal B_L \cup \mathcal B_R, \ \tilde{\alpha}_{n}, \tilde{\beta}_{n} \in \mathcal B_C \ .
\end{equation}
Adding the generator $\sigma$ one gets a group generated by $\delta, \gamma$ and $\sigma$ given by the relations
\begin{equation}
\sigma^4 = 1; \quad \sigma \cdot \gamma \cdot \sigma^{-1} = \gamma^{-1}; \quad \delta \cdot \sigma \cdot \delta^{-1} \cdot \sigma^{-1} = \gamma \ .
\end{equation}

One sees that the relations between the two polygons is not such that one can consider the union of the two polygons as one single polygon with a single dihedral symmetry group acting on it. This is also reflected in the Cartan matrix of the BKM Lie superalgebra, where the inner product between the two chambers occurs with a gradation. We will shortly come to discuss the algebra $\mathcal{ \widetilde G}_5$, where will we see the above mentioned facts.

\subsection{The BKM Lie superalgebra $\mathcal{\widetilde G}_5$ and walls of its Weyl chambers}\label{WallWeylChamber}

Here we put together everything about the algebra $\mathcal{\widetilde G}_5$ in the context of studying the walls of its Weyl chambers. We will use the correspondence with the walls of marginal stability, and the labeling introduced in the previous section to write down the Cartan matrix for the algebra and also study its properties. 

Cheng and Verlinde\cite{Cheng:2008fc} had observed that the walls of marginal stability for the $\tfrac14$-BPS states in the $N=1$ model had a correspondence with the walls of the Weyl chamber of the BKM Lie superalgebra $\mathcal G_1$. Subsequently, Cheng and Dabholkar and Govindarajan and Krishna have shown that for the $N=1,2,3$\cite{Cheng:2008kt} and $4$\cite{Govindarajan:2009qt} cases the fundamental domain for the walls of marginal stability of the $\tfrac14$-BPS states correspond to the Weyl chambers of a family of rank-three BKM Lie superalgebras. Each wall (edge) of the fundamental domain is identified with a real simple root of the BKM Lie superalgebra. Recall that each wall corresponds to a pair of rational numbers $(\tfrac{b}a,\tfrac{d}c)$ which are the intercepts of the wall on the Re($\lambda$)-axis. This is related to a real simple root $\alpha$ of the BKM Lie superalgebra as:
\begin{equation}
(\tfrac{b}a,\tfrac{d}c) \leftrightarrow \begin{pmatrix}a & b \\ c & d \end{pmatrix} \leftrightarrow
\alpha=\begin{pmatrix} 2bd & ad + bc \\ ad + bc & 2 ac \end{pmatrix} \ ,
\end{equation}
with $ac\in N\mathbb{Z}$ and $ad,bc,bd\in \mathbb{Z}$. The norm of the root is\cite{Cheng:2008fc} 
$$
-2\det(\alpha)=2(ad-bc)^2=2\ .
$$ 
The Cartan matrix, $A^{(N)}$, is generated by the matrix of inner products among all real simple roots. The Cartan matrices for the $N=1,2,3$ and $4$ models are given in \eqref{CartanN1}, \eqref{CartanN2}, \eqref{CartanN3} and \eqref{CartanN4}.

Coming to the case of $\mathcal{\widetilde G}_5$, this correspondence goes through even though the number of real simple roots are infinite in number and there are two chambers in the fundamental domain for $N=5$. To construct the Cartan matrix of $\mathcal{\widetilde G}_5$, let us order the real simple roots in the chamber $\mathcal B_L \cup \mathcal B_R$ into an infinite dimensional vector
$$
\mathbf{X}=(\ldots, x_{-2}, x_{-1}, x_{0}, x_{1}, x_2, x_3,\ldots)=( \ldots, \alpha_1, \beta_{-1}, \alpha_0, \beta_0, \alpha_{-1}, \beta_1,\ldots)$$
and similarly the real simple roots in the chamber $\mathcal B_C$ into an infinite dimensional vector as 
$$
\mathbf{\widetilde{X}}=(\ldots, \tilde{x}_{-2}, \tilde{x}_{-1}, \tilde{x}_{0}, \tilde{x}_{1}, \tilde{x}_2, \tilde{x}_3,\ldots)=( \ldots, \tilde{\alpha}_1, \tilde{\beta}_{-1}, \tilde{\alpha}_0, \tilde{\beta}_0, \tilde{\alpha}_{-1}, \tilde{\beta}_1,\ldots).$$

This is precisely the labeling order we introduced on the walls in the previous section. 
Equivalently, let
\begin{equation}
x_m = \left\{\begin{array}{ll} \alpha_{-m/2} \textrm{ or } \tilde{\alpha}_{-m/2}\ , & m\in 2\mathbb{Z}\\
                                            \beta_{(m-1)/2} \textrm{ or } \tilde{\beta}_{(m-1)/2}\ , & m \in 2\mathbb{Z}+1\ .
                                            \end{array} \right.
\end{equation}

The Cartan matrix is given by the matrix of inner products $a_{mn}\equiv \langle x_n , x_m \rangle$ and is given by the infinite dimensional matrix:
\begin{equation}\label{infiniteCartan}
A^{(5)} = (a_{nm})\quad \textrm{where}\quad 
a_{nm}= - 4 \Big( \Lambda_1^{-n} \Lambda_2^{-m} + \Lambda_1^{-m} \Lambda_2^{-n} \Big) + (-1)^{d(x_m) + d(x_n)} 10,
\end{equation}
where $m,n\in \mathbb{Z}$, $\Lambda_1$ and $\Lambda_2$ are the roots of the equation $r^2 - 3 \cdot r + 1 = 0$ and $d(x_m) = 0$ if $x_m \in \mathcal B_L \cup \mathcal B_R$ and $d(x_m) = 1$ if $x_m \in \mathcal B_C$ is the grading on the two sets of roots of the algebra. 

As mentioned before, one can see that the inner products between any two roots in $X$ is equal to the inner product between the corresponding two roots in $\widetilde{X}$. Thus, $X$ and $\widetilde X$ are two copies of the same set of roots with the same inner product matrices. However, a peculiar thing occurs in taking the inner product of the real simple roots with the Weyl vector $\rho$. The inner product of the roots with the Weyl vector $\rho$ satisfies
\begin{equation}
\langle \rho, x_m\rangle =-1\ ,\ \forall \ x_m \in \mathcal{B}_L \cup \mathcal{B}_R\ \quad \textrm{and} \quad \langle \rho, x_m\rangle = 1\ ,\ \forall \ x_m \in \mathcal{B}_C\ .
\end{equation}
\noindent As seen from the above equation, the real simple roots in the chamber $\mathcal B_C$ seem to have the wrong sign when one takes their inner product with the Weyl vector while the real simple roots in the chamber $\mathcal{B}_L \cup \mathcal{B}_R$ have the correct sign. At this point, no explanation for the above fact is known by the author.

\subsubsection*{$D^{(2)}_\infty$-invariance of $\widetilde{\Delta}_{1}(\mathbf{Z})$}

It remains to be proven that $\widetilde{\Delta}_1 (\mathbf{Z})$ gives rise to the denominator identity for the BKM Lie superalgebra $\widetilde{\mathcal G}_5$. One needs to show that $\widetilde{\Delta}_1(\mathbf{Z})$ contains all the real simple roots that one expects from the study of the walls of marginal stability. The $D^{(2)}_\infty$-generators $\gamma$ and $\delta$ act on the roots $x_m$ written as a $2\times2$ matrix as follows:
\begin{align}
\gamma:\  &x_m \longrightarrow \begin{pmatrix} 1 & -1 \\ 5 & -4 \end{pmatrix}\cdot x_m \cdot \begin{pmatrix} 1 & -1 \\ 4 & -3 \end{pmatrix}^{\textrm{T}}\ ,\\
\delta:\  &x_m \longrightarrow \begin{pmatrix} -1 & 1 \\ 0 & 1 \end{pmatrix}\cdot x_m \cdot \begin{pmatrix} -1 & 1 \\ 0 & 1 \end{pmatrix}^{\textrm{T}}\ .
\end{align}
The matrix $\gamma$ is denoted  by $\gamma^{(5)}$ in \cite{Cheng:2008kt}. Under the level $5$ subgroup $G_0(5)\in Sp(2,\BZ)$, the modular form $\widetilde{\Phi}_{2}(\mathbf{Z})$ transforms as\cite{Jatkar:2005bh} 
\begin{equation}
\label{Phitransformation}
\widetilde{\Phi}_{2}(M \cdot \mathbf{Z}) = \{\textrm{det}(CZ+D)\}^2 \widetilde{\Phi}_{2}(\mathbf{Z}),
\end{equation}
where 
\[
M = \begin{pmatrix} A & B \\ C & D \end{pmatrix} \in \textrm{Sp}(2, \mathbb Z), \ M \cdot \mathbf{Z} = (A\mathbf{Z} + B)(C\mathbf{Z}+D)^{-1}, \textrm{ and } C = 0 \textrm{ mod }5.
\]
Consider the subgroup of $G_0(5)$ given by $B = C =0$ and $A^T = D^{-1}$. Under this subgroup, eq. \eqref{Phitransformation} becomes
\begin{equation}
\widetilde{\Phi}_{2}(D^T \cdot \mathbf{Z} \cdot D) = (\textrm{det}D)^2 \widetilde{\Phi}_{2}(\mathbf{Z}) \ .
\end{equation}
Choosing $D = \gamma = \begin{pmatrix} 1 & -1 \\ 5 & -4 \end{pmatrix}$, one sees that $\widetilde{\Phi}_{2}(\mathbf{Z})$ is invariant. Similarly, when $D = \delta = \begin{pmatrix} -1 & 1 \\ 0 & 1 \end{pmatrix}$, or $D = \sigma = \begin{pmatrix} 2 & -1 \\ 5 & -2 \end{pmatrix}$, the modular form $\widetilde{\Phi}_{2}(\mathbf{Z})$ is invariant. Thus, we see that the modular form $\widetilde{\Phi}_{2}(\mathbf{Z})$ is invariant under the action of $\gamma$, $\delta$, and $\sigma$, which means that under the action of $\gamma$, $\delta$ and $\sigma$,
\begin{equation}
\widetilde{\Delta}_1 (\mathbf{Z}) \rightarrow \pm \widetilde{\Delta}_1 (\mathbf{Z}) \ .
\end{equation}
One can show that the sign must be $+1$ by observing that any pair of terms in the Fourier expansion of $\widetilde{\Delta}_1 (\mathbf{Z})$ related by the action of $\gamma$ (resp. $\delta, \sigma$) appear with the same Fourier coefficient. For instance, the terms associated with the two simple roots $\alpha_0$ and $\beta_0$ related by the action of $\delta$ appear with coefficient $+1$. Similarly, the terms associated with the real simple roots $\beta_0$ and $\beta_{-1}$ related by a $\gamma$-translation also appear with coefficient $+1$. Thus, we see that $\widetilde{\Delta}_1 (\mathbf{Z})$ is invariant under the full dihedral group generated by $\delta, \gamma$ and $\sigma$. This provides an \textit{all-orders} proof that the infinite real simple roots given by the vector $\mathbf{X}$ all appear in the Fourier expansion of $\widetilde{\Delta}_1 (\mathbf{Z})$.

The $q\rightarrow s^5$ symmetry of the modular form is equivalent to the symmetry generated by the dihedral generator, $y$, as defined in Eq. \eqref{Dinfinity}.

\section{Walls of marginal stability and recurrence}

Here we study the the general structure of the walls of marginal stability for the CHL models. In\cite{Cheng:2008kt} Cheng and Dabholkar gave an arithmetic argument underlying the walls of marginal stability for the $N=1,2$ and $3$ models. They observed that the intercepts of the walls in the fundamental chamber are given by the rational numbers in the Stern-Brocot tree which is formed by taking the median, $\tfrac{b+d}{a+c}$, of the successive pair of rationals $\{ \tfrac{b}{a} , \tfrac{d}{c}\}$ starting from $\tfrac{\pm 1}{0}, \tfrac01$. Each of the successive rows gives the intercepts of the walls of marginal stability for the $N=1,2$ and $3$ models. However, such a structure, which also respects the consistency conditions imposed by the orbifolding (namely that the neighboring rational numbers have the product of their denominators divisible by $N$), does not exist for $N>3$. Also, from Sen's analysis, an infinite number of walls are expected for $N>3$, but the Stern-Brocot tree gives only a finite number of rationals at any level. Thus, it would appear that either an arithmetic structure does not exist for $N>3$, or if one exists it is given by a different kind of series (in place of the Stern-Brocot series) for $N\geq 4$. 

We show that an arithmetic structure exits for $N>3$ and use it to compute the walls of marginal stability for the $N=4,5$ and $6$\footnote{The following analysis also holds for the $N=7$ case, but only for the chamber $\mathcal{B}_L \cup \mathcal{B}_R$. The intercepts in the chamber $\mathcal B_C$ is a bit more complicated and the analysis does not give the intercepts in this chamber.} models. All these models have an infinte number of walls in the fundamental domain. The walls of marginal stability for these models are generated by a pair of linear recurrance relations. We will see that there are many universal properties which are given only as a function of $N$ but the form of these functions across different $N$ remains the same. One starts with the observation that in each chamber the intercepts on the Re($\lambda$)-axis are given by two sequences of the form
\begin{equation}\label{Anrec}
A_n = (N-2) A_{n-1} - A_{n-2} ,
\end{equation}
and 
\begin{equation}\label{Bnrec}
B_n = (N-2) B_{n-1} - B_{n-2} ,
\end{equation}
where $N=4, 5$ and $6$. As can be seen, the form of the sequences is the same for all $N$. One can compute the above sequences by finding solutions to the recurrence equations and using two initial values to fix the sequence. They can also be written down from a generating function as is explained later\footnote{Sen generates all the walls by the action of the matrix $g_0 \equiv \begin{pmatrix}1-N & 1 \\ -N & 1 \end{pmatrix}$, which translates the walls by an even number of steps. Repeated action of $g_0$ generates all the infinite number of walls.}. The \textit{characteristic equation} (equation satisfied by the solution to the ansatz $A_n = \ell^n$) is
\begin{equation}
\label{Characteristicequation}
p(\ell) \equiv \ell^2 - (N-2) \cdot \ell =-1 ,
\end{equation}
which can also be seen to be of the same form for all $N$.
The solutions to the linear recurrance depend on the nature of the roots of the characteristic equation. For identical roots, as one has in the case of $N=4$, one has the general solution
\begin{equation}
A_n = C \Lambda^n + D n \Lambda^n\ ,
\end{equation}
while for distinct roots, as is the case for $N=5$ and $6$, one has
\begin{equation}
A_n = C \Lambda_1^n + D \Lambda_2^n\ ,
\end{equation}
where $C$ and $D$ are fixed from two initial conditions.

The walls in the chambers $\mathcal B_L \cup \mathcal B_R$ and $\mathcal B_C$ are given by intercepts which come from linear recurrance sequences like the above. Let us study each case separately, before putting together a general picture. 

\section*{The walls of marginal stability for $N=4$}

The characteristic equation and the (identical) roots for the walls of marginal stability for the $N=4$ model are
\begin{displaymath}
p_4(\ell) \equiv \ell^2 - 2 \cdot \ell +1 = 0, \quad \textrm{ with roots } \Lambda = 1 \ .
\end{displaymath}
Using initial conditions to determine the sequence, we have for the series $A_n$ and $B_n$
\begin{equation}\label{seriesN4}
A_n = n ;\quad B_n = 2n+1 \ .
\end{equation}
Using the above sequence one finds that the intercepts on the Re($\lambda$)-axis in the two chambers $\mathcal B_L \cup \mathcal B_R$ are given by 
\begin{equation}\label{lchamber4}
\mathcal B_L \cup \mathcal B_R = \big \{ \frac{A_n}{B_n} , \frac{B_n}{4 \cdot A_{n+1}} \big \} , n \in \mathbb{Z} ,
\end{equation}
One can check that the limit points for both the sequences in \eqref{lchamber4} are 
\begin{equation}
\lim_{n \rightarrow \infty} \frac{A_n}{B_n} = \lim_{n \rightarrow \infty} \frac{B_n}{4 \cdot A_{n+1}} = \tfrac12
\end{equation}
in keeping with the fact that the point $\tfrac12$ is the limit point for walls, starting at $0$ and $1$, in the fundamental chamber. 


As before, it is again convenient to divide the intercepts into two sequences $\alpha_n$ and $\beta_n$ and use a labeling which will help in studying them and also writing the Cartan matrix of the BKM Lie superalgebra. In the notation of \cite{Govindarajan:2009qt} the roots of the BKM Lie superalgebra are given from the intercepts of the corresponding walls by
\begin{eqnarray}\label{rootsN4}
\alpha_n \equiv \Big \{\frac{B_{n-1}}{4 \cdot A_n} , \frac{A_{n}}{B_{n}} \Big \} \leftrightarrow \begin{pmatrix} 2 \cdot A_{n} B_{n-1} & 4 \cdot A_n^2 + B_{n}B_{n-1} \\ 4 \cdot A_n^2 + B_{n}B_{n-1} & 8 \cdot A_n B_{n}\end{pmatrix} \nonumber \\
\beta_n \equiv \Big \{\frac{A_{n+1}}{B_{n}} , \frac{B_{n}}{4 \cdot A_{n}} \Big \} \leftrightarrow \begin{pmatrix} 2 \cdot A_{n+1} B_{n} & 4 \cdot A_{n}A_{n+1} + B_n^2 \\ 4 \cdot A_{n}A_{n+1} + B_n^2 & 8 \cdot A_n B_{n}\end{pmatrix} 
\end{eqnarray}

Using the correspondence between the walls of marginal stability and the roots, one can construct the Cartan matrix for the BKM Lie superalgebra, $\widetilde{\mathcal G}_4$, by taking the inner product between the roots using their form given in \eqref{rootsN4} and using \eqref{seriesN4} . The Cartan matrix is given by \eqref{CartanN4}. This is indeed the same Cartan matrix as obtained by constructing the BKM Lie superalgebra directly from the modular forms via the Weyl-Kac-Borcherds denominator formula. One could use this to write down the Cartan matrices of the BKM Lie superalgebras of the models where direct computation of the modular forms is difficult.

The $\Gamma_0(4)$ element $\gamma^{(4)} = \begin{pmatrix} 1 & -1 \\ 4 & -3\end{pmatrix}$ acts as a translation on the intercepts taking $\alpha_n \mapsto \alpha_{n+1}$ and $\beta_n \mapsto \beta_{n-1}$, while the element $\delta = \begin{pmatrix} -1 & 1 \\ 0 & 1\end{pmatrix}$ acts as a reflection on the intercepts exchanging the walls $\delta: \alpha_n \leftrightarrow \beta_n$. The two elements $(\gamma^{(4)}, \delta)$ form a dihedral group, which is the symmetry group of the polygon formed by the walls in the fundamental chamber. 

\section*{Walls of marginal stability for $N=6$}

The characteristic equation for the $N=6$ case is given by
\begin{displaymath}
\label{chareq6}
p_6(\ell) \equiv \ell^2 - 4 \cdot \ell +1 = 0 \ .
\end{displaymath}
Using initial conditions to determine the sequences, the intercepts of the walls of marginal stability in the $\mathcal B_L \cup \mathcal B_R$ chambers are given by 
\begin{equation}\label{lchamber6}
\mathcal B_L \cup \mathcal B_R = \big \{ \frac{A_n}{B_n} , \frac{B_n}{6 \cdot A_{n+1}} \big \} , n \in \mathbb{Z} ,
\end{equation}
where the two sequences $A_n$ and $B_n$ are given by
\begin{equation}
A_n = \frac{1}{2\sqrt 3} \Big( \Lambda_1^n - \Lambda_2^n \Big), \ B_n = \frac{1+\sqrt 3}{2} \Lambda_1^n + \frac{1-\sqrt 3}{2} \Lambda_2^n \ .
\end{equation}
The intercepts of the walls in the domain $\mathcal B_C$ are given by 
\begin{equation}\label{mchamber6}
\mathcal B_C = \big \{ \frac{\widetilde{A}_n}{2 \cdot \widetilde{B}_n} , \frac{\widetilde{B}_{n+1}}{3 \cdot \widetilde{A}_n} \big \} ,
\end{equation}
where the two sequences $\widetilde{A}_n, \widetilde{B}_n$ are given as
\begin{equation}
\widetilde{A}_n = \frac{\sqrt 3 - 1}{2 \sqrt 3} \Lambda_1^n + \frac{\sqrt 3 + 1}{2 \sqrt 3} \Lambda_2^n, \ \widetilde{B}_n = \frac{2-\sqrt 3}{2} \Lambda_1^n + \frac{2+\sqrt 3}{2} \Lambda_2^n,
\end{equation}
where in all the equations above $\Lambda_1 = 2 + \sqrt 3$ and $\Lambda_2 = 2 - \sqrt 3$, which are nothing but the roots of the characteristic equation \eqref{chareq6}. One can check that the limit points for the ratios of the sequences in \eqref{lchamber6} and \eqref{mchamber6} are $\tfrac12 (1 \pm \sqrt{\tfrac13})$ which is consistent with Sen's analysis\cite{Sen:2007vb}. 

As before, choosing a labeling in the chambers $\mathcal B_L \cup \mathcal B_R$ we form two sequences given as
\begin{equation}
\alpha_n = \begin{pmatrix} B_{n-1} & A_n \\ 6 \cdot A_{n} & B_{n} \end{pmatrix} \ \textrm{ and } \ \beta_n = \begin{pmatrix}  A_{n+1} & B_{n} \\ B_{n} & 6 \cdot A_{n} \end{pmatrix}
\end{equation}
and in the chamber $\mathcal B_C$ as
\begin{equation}
\widetilde{\alpha}_n = \begin{pmatrix} \widetilde{B}_{-n+1} & 3 \widetilde{A}_{-n} \\ \widetilde{A}_{-n+1} & 2 \widetilde{B}_{-n+1} \end{pmatrix} \ \textrm{ and } \
\widetilde{\beta}_n = \begin{pmatrix} \widetilde{A}_{n+1} & 2 \widetilde{B}_{n+1} \\ \widetilde{B}_{n+2} & 3 \widetilde{A}_{n+2} \end{pmatrix}\ .
\end{equation}

Using the correspondence between the walls of marginal stability and the roots of the BKM Lie superalgebra, one can construct the Cartan matrix for the $N=6$ model just as one did for the $N=4$ and $5$ models. The Cartan matrix for the BKM Lie superalgebra, $\widetilde{\mathcal G}_6$, corresponding to the $N=6$ model is given by
\begin{equation}\label{CartanN6}
A^{(6)} = (a_{nm})\quad \textrm{where}\quad 
a_{nm}= - 2 \Big( \Lambda_1^{-n} \Lambda_2^{-m} + \Lambda_1^{-m} \Lambda_2^{-n} \Big) + (-1)^{d(x_m) + d(x_n)} 6,
\end{equation}
with $m,n\in \mathbb{Z}$, $\Lambda_1, \Lambda_2$ are the roots of the equation $r^2 - 4 \cdot r + 1 = 0$, and $d(x_m) = 0$ if $x_m \in \mathcal B_L \cup \mathcal B_R$ and $d(x_m) = 1$ if $x_m \in \mathcal B_C$ is the grading on the two sets of roots of the algebra. 

The $\Gamma_0(6)$ element $\gamma^{(6)} = \begin{pmatrix} 1 & -1 \\ 6 & -5\end{pmatrix}$ acts as a translation on the intercepts taking $\alpha_n \mapsto \alpha_{n+1}$ and $\beta_n \mapsto \beta_{n-1}$, while the element $\delta = \begin{pmatrix} -1 & 1 \\ 0 & 1\end{pmatrix}$ acts as a reflection on the intercepts exchanging the walls $\delta: \alpha_n \leftrightarrow \beta_n$. The two elements $(\gamma^{(6)}, \delta)$ form a dihedral group for the polygon generated by the roots. 

\subsection*{General structure of the walls}
Now we are ready to put together a general structure for the walls of marginal stability in the case of general $N$. As one can see from the above analysis, the general structure of the walls has a pattern which is same across $N$. For all $N$, the walls are generated by two sequences, $A_n, B_n$, for both the chambers $\mathcal B_L \cup \mathcal B_R$ and $\mathcal B_C$. The sequences are generated by linear recurrance relations given by the respective characteristic equations \eqref{Characteristicequation}. The form of this equation is the same for all the $N$. The sequences $A_n, B_n$ in the chambers $\mathcal B_L \cup \mathcal B_R$ are generated by the generating functions
\begin{equation}
\phi(x) = \frac{x}{p(x)}, \ \textrm{ and } \ \chi(x) = \frac{1+x}{p(x)},
\end{equation}
where $p(x)$ is the characteristic equation given in \eqref{Characteristicequation}. The above equations depend on $N$ in the denominator $p(x)$, but their general form (and even the general form of the characteristic equation \eqref{Characteristicequation}) remains the same for $N=4,5,6$. 

The intercepts in the chambers $\mathcal B_L \cup \mathcal B_R$ are given by the general form 
\begin{equation}\label{lchamber}
\mathcal B_L \cup \mathcal B_R = \big \{ \frac{A_n}{B_n} , \frac{B_n}{N \cdot A_{n+1}} \big \} , n \in \mathbb{Z} ,
\end{equation}
for all $N$. The intercepts in the chamber $\mathcal B_C$ are given by the general form 
\begin{equation}
\mathcal B_C = \big \{ \frac{\widetilde{A}_n}{P \cdot \widetilde{B}_n} , \frac{\widetilde{B}_{n+1}}{Q \cdot \widetilde{A}_n} \big \} ,
\end{equation}
where $P, Q$ are the prime factors of $N (5 = 5\cdot 1, 6 = 2 \cdot 3)$. One also sees that the form of the sequences, $\alpha_n$ and $\beta_n$, denoting the semi-circles in $\mathcal B_L$ and $\mathcal B_R$ with intercepts given by 
\begin{equation}
\alpha_n = \begin{pmatrix} B_{n-1} & A_n \\ N \cdot A_{n} & B_{n} \end{pmatrix} \ \textrm{ and } \ \beta_n = \begin{pmatrix}  A_{n+1} & B_{n} \\ B_{n} & N \cdot A_{n} \end{pmatrix} \ .
\end{equation}
is the same for all $N$. Similarly, the sequences, $\widetilde{\alpha}_n$ and $\widetilde{\beta}_n$, denoting the semi-circles in the chamber $\mathcal B_C$ given by
\begin{equation}
\widetilde{\alpha}_n = \begin{pmatrix} \widetilde{B}_{-n+1} & Q \cdot \widetilde{A}_{-n} \\ \widetilde{A}_{-n+1} & P \cdot \widetilde{B}_{-n+1} \end{pmatrix} \ \textrm{ and } \ \widetilde{\beta}_n = \begin{pmatrix} \widetilde{A}_{n+1} & P \cdot \widetilde{B}_{n+1} \\ \widetilde{B}_{n+2} & Q \cdot \widetilde{A}_{n+2} \end{pmatrix}
\end{equation}
is the same for the $N=5, 6$ which have a central chamber. The limit points for the ratios appearing in the above sequences are also given only as a function of $N$. 

Also, the form of the transformations which translate the intercepts, namely the $\Gamma_0(N)$ element $\gamma^{(N)} = \begin{pmatrix} 1 & -1 \\ N & 1-N\end{pmatrix}$ acting as a transformation on the intercepts taking $\alpha_n \mapsto \alpha_{n+1}$ and $\beta_n \mapsto \beta_{n-1}$ (and similarly $\widetilde{\alpha}_n \mapsto \widetilde{\alpha}_{n+1}$ and $\widetilde{\beta}_n \mapsto \widetilde{\beta}_{n-1}$) is the same across $N =4,5,6$. The transformation that exchanges the intercepts, $\delta: \alpha_n \leftrightarrow \beta_n$ in the chambers $\mathcal B_L \cup \mathcal B_R$ and $\delta: \widetilde{\alpha}_n \leftrightarrow \widetilde{\beta}_n$ in $\mathcal B_C$, remains the same too. 

Thus, we see that the general form of the walls of marginal stability for the models $N \geq 4$ follows a general pattern which is same for all the $N$. As mentioned before, this is similar to the arithmetic structure found by Cheng and Dabholkar, and extends it to $N \geq 4$. 

\section*{Walls of marginal stability for $N=7$}
Before we conclude this section, let us briefly also analyse the situation for the $N=7$ case. The intercepts of the walls in the chamber $\mathcal B_L \cup \mathcal B_R$ for $N=7$ is given, as it is for $N=4,5$ and 6, by recurrence relation of the form \eqref{Anrec} and \eqref{Bnrec}. The two sets of walls, $\alpha_n$ and $\beta_n$, in the  $\mathcal B_L \cup \mathcal B_R$ chamber are given by 
\begin{equation}\label{lchamberN7}
\mathcal B_L \cup \mathcal B_R = \big \{ \frac{A_n}{B_n} , \frac{B_n}{7 \cdot A_{n+1}} \big \} , n \in \mathbb{Z} ,
\end{equation}
where the two sequences $A_n$ and $B_n$ are given by
\begin{equation}
A_n = \frac{1}{\sqrt{21}} \Big( \Lambda_1^n - \Lambda_2^n \Big), \ B_n = \frac{\sqrt{21} + 7}{2\sqrt{21}} \Lambda_1^n + \frac{\sqrt{21}- 7}{2\sqrt{21}} \Lambda_2^n \ .
\end{equation}

The limit points for the intercepts in the chamber, $\mathcal B_L \cup \mathcal B_R$ are given by $\tfrac12 (1 \pm \sqrt{1 - \tfrac47})$. As before, the element $\gamma \equiv \begin{pmatrix}1 & -1 \\ 7 & -6 \end{pmatrix}$, acts as a transformation on the intercepts taking $\alpha_n \mapsto \alpha_{n+1}$ and $\beta_n \mapsto \beta_{n-1}$ (and similarly $\widetilde{\alpha}_n \mapsto \widetilde{\alpha}_{n+1}$ and $\widetilde{\beta}_n \mapsto \widetilde{\beta}_{n-1}$). 

However, the intercepts of the walls in the central chamber are more complicated and are not given by the above analysis.

\section{Discussion}

In the previous sections we constructed the BKM Lie superalgebra, $\mathcal{\widetilde G}_5$, and also studied the walls of the Weyl chamber of the algebra in relation to the walls of marginal stability of the $\tfrac14$-BPS states. While this algebra follows the general pattern of the algebras for $N<5$, some peculiar aspects have not been understood completely and this discussion is aimed at underlining them. First, is the fact that the real simple roots from the chamber $\mathcal B_C$ do not have the right sign in their inner product with respect to the Weyl vector of the algebra. The roots in both the chambers are exact one-to-one copies of each other, including, having the same inner product between two corresponding pair of roots in either chamber and only differ in their inner product with $\rho$. This leads to the possibility that each of the set of roots forms an algebra by itself. The same also occurs for the roots of $\mathcal{\widetilde G}_6$, which has two chambers in the fundamental domain.

Secondly, in Borcherd's construction of holomorphic infinite products, the BKM Lie superalgebras were related to the modular forms themselves, where as in all the above discussed examples of the CHL strings it is the square root of the modular forms which lead to a BKM Lie superalgebra and not the modular forms themselves. The occurrence of the square root is not understood. 

Finally, the walls of marginal stability of $N=7$ seems to differ from the $N=5$ and $6$ which also have two chambers in the fundamental domain. While the roots in the chamber $\mathcal B_L \cup \mathcal B_R$ seem to be along the same lines as the analysis for $N=5$ and $6$, the intercepts in the central chamber are not given by the same method.

\section{Conclusion}

In this paper we have studied the $\mathbb Z_5$-orbifolded CHL string in some detail, starting from the modular forms generating the degeneracy of the half and quarter BPS states in the theory. After explicitly constructing the genus-two Siegel modular form generating the $\tfrac14$-BPS degeneracies, we see that  the modular forms generated from the additive and multiplicative lifts do not match for the $N=5$ case. Taking the product form to be the proper expansion, we constructed the BKM Lie superalgebra corresponding to the Siegel modular form from the Weyl-Kac-Borcherds denominator identity. This BKM Lie superalgebra, like the $N=4$ case, has an infinite number of real simple roots. We also studied the walls of marginal stability for the $N=4,5$ and $6$ cases in general and found that they are generated by linear recurrance relations. The correspondence between the Weyl chambers of the BKM Lie superalgebra and the walls of marginal stability of the $\tfrac14$-BPS states continues to hold for the $N=5$ case also. Based on this, we also propose a Cartan matrix for the BKM Lie superalgebra $\mathcal{\widetilde G}_6$ for the $\mathbb Z_6$-orbifolded CHL model.

\section*{Acknowledgements}

I would like to thank Prof. Suresh Govindarajan for numerous helpful discussions. I also thank Prof. Nils Scheithauer and Prof. Dileep Jatkar for clarifications and discussions. I am very grateful to Prof. Ashoke Sen, and Prof. Urmie Ray for their kind support. Most of this work was done during my stay at HRI, and I would like to thank the staff and members of HRI for all the support extended during my stay.

\bibliography{master}
\end{document}